\documentstyle[aps,prl,twocolumn,epsf]{revtex}
\begin{document}
\draft
\title{Probing strange stars and color superconductivity by 
$r$-mode instabilities in millisecond pulsars}
\author{Jes Madsen}
\address{Institute of Physics and Astronomy, University of Aarhus, 
DK-8000 \AA rhus C, Denmark}
\date{December 20, 1999; scheduled for Physical Review Letters July 3, 2000}
\maketitle

\begin{abstract}
$R$-mode instabilities 
in rapidly rotating quark matter stars (strange
stars) lead to specific signatures in the evolution of pulsars with
periods below 2.5 msec, and may explain the apparent lack of very
rapid pulsars. Existing data seem consistent with pulsars being strange
stars with a normal quark matter phase surrounded by an insulating
nuclear crust. In contrast, quark stars 
in a color-flavor-locked (CFL) phase are ruled out.
Two-flavor color superconductivity (2SC) is
marginally inconsistent with pulsar data. 
\end{abstract}


Starting with Andersson's realization, that rotating relativistic stars
are generically unstable against the $r$(otational)-mode instability
\cite{and98}, a
series of papers have investigated the many implications for
gravitational radiation detection and the evolution of pulsars
\cite{frimor98,linowe98,oweal98,mad98,koj98,ankosc99,ankost99,lev99,locfri99,kokste99,limeow99,kojhos99,rezmaa99,rezjah99,linmen99,yosal99,bilush99,relash99,holai99}. 
Originally it appeared that young, hot neutron stars would spin
down to rotation periods of order 10 msec within their first year of
existence. In contrast, thousands of years would be
required for any $r$-mode driven spin-down of a hot strange star or a
neutron star with significant quark matter content, and the rotation
period would not increase above 3 msec in this case, making young,
rapid pulsars potential ``smoking guns'' for quark matter
(meta)stability \cite{mad98}. 
Decisive for the localization of the instability
regimes were the viscosities damping the modes, with strange matter
characterized by a huge bulk viscosity relative to nuclear matter.

Recently Bildsten and Ushomirsky \cite{bilush99} pointed out, that
a very important effect damping $r$-modes in neutron stars had been
overlooked. This is damping due to viscosity in the boundary layer
between the oscillating fluid and the nearly static crust, which is more
than $10^5$ times stronger than that from the shear in the interior.
Matching of boundary conditions at the crust is particularly important
for $r$-modes, since these are characterized by significant horizontal
flows. As a result, $r$-modes in neutron stars are only important for
rotation periods faster than 2 msec, and only for very high core
temperatures (with the possible exception of the very brief time span
before the crust forms). Unless it turns out, that neutron stars are
able to spin down significantly before their crust solidifies, this
means that young, rapid pulsars could be neutron stars after all; not
just strange stars.

But other, perhaps even better, probes of strange stars result from
the $r$-mode instability as demonstrated below. 
Furthermore, pulsar data turn out to be very
sensitive probes of color superconductivity in quark matter
\cite{raj99}. These prospects are pursued in the following.

In contrast to neutron stars, the quark matter fluid in a strange star
need not be at rest at the base of the
crust, and therefore the $r$-modes in a quark matter star are not damped
significantly
by ``surface rubbing''. If strange quark matter is absolutely stable
(having lower energy per baryon than nuclear
matter), strange stars may be bare, consisting of quark matter fluid 
all the way to the surface. In this case, no surface rubbing takes place.
But even in the more likely case, where gas from
the supernova explosion or later accretion reaches the surface, the
crust formed floats on top of a huge electrostatic potential,
separated from the quark surface. This is a result of
the strong interactions confining the quarks much tighter than
the electrostatic forces holding the electrons, so that electrons create
an atmosphere of a few hundred Fermi thickness \cite{alfaol86}, effectively
separating quark matter from the nuclear crust. Some viscosity results
from the interaction between the outer part of the electron atmosphere
and the base of the crust, but as illustrated later, due to the
low density in the strange star crust relative to that of a neutron
star, this effect
is only dominant when other sources of viscosity are exponentially
suppressed in the case of a color-flavor locked quark phase.

Rapidly rotating millisecond pulsars (periods below 2.5--3 msec)
are unstable to the
$r$-mode instability for core temperatures in the range of
${\rm a~few}\times 10^{5}$--${\rm a~few}\times 10^{7}$K, if they are
quark stars with a normal fluid quark phase. Interestingly,
the fastest pulsars known may be just outside this
window of instability, reaching it within $10^4$ years due to cooling.
When a pulsar reaches the instability window it will slow down by
gravitational wave emission on a time scale of $10^4$--$10^5$ years
to a period near 2.5 msec, slowing down further on a
much longer time scale due to magnetic
dipole braking. As demonstrated below,
the $r$-mode spin-down would be characterized by
a so-called braking index with an unusually high value of $N\approx 9$, a
clear observational indication of the process.
Some pile-up of pulsar periods near 3 msec, and an underrepresentation
of short periods would be expected in this scenario, apparently
consistent with the data.

In contrast, quark matter with diquark pairing into a
color-flavor locked phase \cite{raj99}, would have an
exponential reduction in the viscosities, expanding the $r$-mode
instability region to encompass low-mass x-ray binaries
(LMXB's) as well as many known pulsars, which should then rapidly spin down,
in disagreement with observations. Unless an as yet unconsidered viscous
effect could prevent this \cite{visc}, it seems that these pulsars cannot be
quark stars with properties expected for CFL. A 2-flavor color
superconducting phase (2SC) \cite{raj99} has less dramatic consequences, 
but still
seems marginally ruled out by the data. Metastable quark
matter in CFL or 2SC-phases cannot be ruled out, however, since a hybrid star
with quark matter in the interior and nuclear matter in the outer
layers, probably with a mixed phase in between, must obey the crust
boundary condition as an ordinary neutron star leading to surface
rubbing.

The critical rotation frequency for a given stellar model as a
function of temperature follows from
\begin{equation}
{{1}\over{\tau_{\rm gw}}}+{{1}\over{\tau_{\rm sv}}}+{{1}\over{\tau_{\rm
bv}}}+{{1}\over{\tau_{\rm sr}}} =0 ,
\label{limit}
\end{equation}
where $\tau_{\rm gw}<0$ is the characteristic time scale for energy loss
due to gravity wave emission, $\tau_{\rm sv}$ and $\tau_{\rm bv}$ are
the damping times due to shear and bulk viscosities, and $\tau_{\rm sr}$
is the surface rubbing time scale. Surface
rubbing is decisive for neutron stars \cite{bilush99}, 
whereas ${{1}/{\tau_{\rm sr}}} =0$  for bare quark stars, and is
suppressed by more than 5 orders of magnitude even for strange stars with
maximal crust.

Ref.\ \cite{mad98} used an analytic description of
$r$-mode instability in uniform stars
\cite{kokste99} to derive the characteristic
time scales for strange stars. A strange star has nearly constant density
except for masses very close to the gravitational instability limit, so
a polytropic equation of state with a low index, $n$, provides a very
good approximation. The case $n=0$ corresponding to constant density was
discussed in \cite{kokste99}, whereas $n=1$ was studied in \cite{limeow99}.
The time scale for gravity wave emission is
\begin{equation}
\tau_{\rm gw}= -3.26(1.57)~{\rm s}\left( {{\pi G \bar\rho}/
{\Omega^2}}\right)^3 ,
\end{equation}
where prefactors outside (inside) parentheses correspond to $n=1$ (0),
$G$ is the gravitational constant, $\Omega$ is the angular rotation
frequency, and $\bar\rho$ is the mean density.
With shear viscosity coefficient taken from \cite{heipet93}, 
the time scale for shear viscous damping is
\begin{equation}
\tau_{\rm sv}= 5.37(2.40)\times 10^8~{\rm s}(\alpha_S/0.1)^{5/3}
T_9^{5/3} .
\end{equation}
Here, $T_9$ denotes temperature in units of $10^9$K, and $\alpha_S$ is
the strong coupling.

The bulk viscosity of strange quark matter
depends mainly on the rate of
$u+d\leftrightarrow s+u$,
which is the fastest of the reactions trying to reestablish weak equilibrium
between massive strange quarks and the much lighter up and down quarks.
To very good approximation the bulk viscosity is given by
\cite{mad92}
$
\zeta = {{\alpha T^2}/[{(\kappa\Omega )^2 +\beta T^4}}],
$
with $\alpha$ and $\beta$ given in \cite{mad92}. 
For the dominant $r$-mode, $\kappa=2/3$.
A low (high) $T$-limit is relevant when the first
(second) term in the denominator dominates.
In cgs-units, the low-$T$ limit is \cite{mad92}
$
\zeta^{\rm low} \approx 3.2\times 10^3 m_{100}^4 
\rho T^2 (\kappa\Omega)^{-2},
$
where $m_{100}$ is the strange quark mass in units of 100~MeV.
The high-$T$ limit takes over for $T> 10^9$K. Here \cite{mad92},
$
\zeta^{\rm high} \approx 3.8\times 10^{61} m_{100}^4 \rho^{-1} T^{-2}.
$
For the bulk viscous damping time the approximation in 
\cite{kokste99} used in \cite{mad98} has turned out to be
too crude, since bulk viscosity
coupling to the $r$-modes happens at second order. Lindblom {\sl et al.}
\cite{limeow99} reevaluated $\tau_{\rm bv}$ for a
strange star in the low-$T$ limit and found 
\begin{equation}
\tau_{\rm bv}^{\rm low}= 0.886~{\rm s}\left( {\pi G \bar\rho}/
{\Omega^2} \right) T_9^{-2} m_{100}^{-4} ,
\end{equation}
The prefactor here is 7 times smaller than used in \cite{mad98}, and
the scaling with $\Omega$ is opposite, resulting in
some changes in the results, though not in the conclusions of
\cite{mad98}. But now also the high-$T$ limit becomes important.
In this limit (not considered in \cite{mad98,limeow99})
\begin{equation}
\tau_{\rm bv}^{\rm high}= 0.268~{\rm s}\left( {\pi G \bar\rho}
/ {\Omega^2} \right)^{2} T_9^2 m_{100}^{-4} .
\end{equation}
Because $\tau_{\rm bv}^{\rm high}$ increases with temperature, it can lead to
$r$-mode instability for very high $T$.

Figure 1 shows the regions of $r$-mode (in)stability in a plot of
pulsar spin frequency ($\nu \equiv \Omega/(2\pi)$) versus temperature
for a strange star with mass $M=1.4M_\odot$ and radius $R=10$km. 
An $n=1$ polytrope was
assumed, but conclusions are not changed for uniform density. 
Also indicated are the positions of LMXB's, presumed to be old
pulsars being spun-up by accretion to eventually become rapid
millisecond pulsars, as well as the positions of the two most rapidly
spinning pulsars known, with periods of 1.5578 and 1.6074 msec ($\nu\approx
642$ and 622 Hz). It should be stressed that
the core temperatures are uncertain upper 
limits derived from x-ray limits on the surface temperatures,
increased by roughly two orders of
magnitude to include effects of an insulating crust. The actual numbers
are taken from \cite{bilush99}, valid for a neutron star model, but
similar limits apply to strange stars with a significant crust. Bare
strange stars or stars with a thin crust 
would have a core temperature close to the surface
temperature, moving them closer to or even inside the region of $r$-mode
instability. Completely bare strange stars are very
poor emitters of radiation below the quark matter plasma frequency of 20
MeV \cite{alfaol86}, but even a
thin crust/atmosphere would allow normal thermal
radiation. For simplicity both categories are denoted ``bare'', but
conclusions based on surface temperatures only
relate to strange stars with a tiny layer of surface pollution.

One notes that the LMXB's are well within the region stable against
$r$-mode instabilities, allowing them to accrete and speed-up unhindered
by the instability. The rapid pulsars are also apparently in the stable
regime (at least for $m_{100}=2$), but as the time scale for cooling to
$10^7$K is only around $10^4$years \cite{schal97}, 
they should soon enter the unstable
region and start spinning down. Since $\tau_{\rm gw}\ll \tau_{\rm cool}$, 
the pulsars should follow a track indistinguishable from the curve
marking the instability region, corresponding to an unusually high
braking index of $N\approx 9$. This value follows because $\Omega\propto
T^{1/2}\propto t^{-1/8}$ (the latter comes from $t\propto \tau_{\rm
cool}\approx 10^{-4}{\rm yr}T_9^{-4}$ for standard neutrino cooling of a
quark star \cite{schal97}). 
Thus $\dot\Omega\propto -{1\over 8}t^{-9/8}$, $\ddot\Omega
\propto {1\over 8}{9\over 8}t^{-17/8}$, and the braking index
$N\equiv \Omega\ddot\Omega/\dot\Omega^2 =9$.
The star will reach a spin frequency of 400Hz
(2.5 msec rotation period) within a cooling time scale of
$10^5$years. Notice that pulsars with the highest spin frequencies need
less time to reach the region of instability. This may explain why no
frequencies above 642Hz have been observed, and the spin-down to 2.5--3 msec
by the $r$-mode instability may lead to some clustering of
observed rotation periods around this value (not inconsistent with data,
but statistics is not overwhelming due to a low number of objects). 
No similar effects arise in
the case of ordinary neutron stars, where $r$-mode instabilities only
seem to work at frequencies above 500Hz, and then mainly for $T\approx
10^{10}$K (dashed curve in Fig.~1).

Notice that the agreement with pulsar data only remains valid if an
insulating crust allows the bulk temperature of the pulsar to be some
two orders of magnitude higher than the observed upper limits on the
surface temperature (about $6\times 10^5$K and $9\times 10^5$K for the
pulsars plotted) to locate the pulsars to the right of the $r$-mode
instability range. A position inside or to the left of this regime seems
ruled out (the latter because pulsars spun up in the LMXB-domain would
have to cross the instability regime before reaching a position to the
left). Therefore, strange stars without a significant crust (having comparable
surface and bulk temperatures) are ruled out as
models for these rapid pulsars unless they are completely bare and
therefore hidden in x-rays.

Superfluidity in the quark phase completely changes the behavior. If
quark pairing is characterized by an energy gap, $\Delta$, reaction
rates involving two quarks (as relevant for
bulk as well as shear viscosities) are suppressed by a factor
$\exp(-2\Delta/T)$, assuming equal behavior for all quark flavors, as
expected in a high density color-flavor locked phase. This
increases the bulk viscous time scale by
$\exp(2\Delta/T)$ and $\tau_{\rm sv}$ (including screening) by
$\exp(\Delta /(3T))$, significantly increasing the parameter space where
the $r$-mode instabilities are active. In fact at low $T$ the viscosity
is now determined by shear due to electron-electron scattering or by
surface rubbing.
The time scale for electron shear is 
$
\tau_{\rm sv}^{ee}\approx 2.95\times 10^{7} {\rm s} (\mu_e/\mu_q)^{-14/3}
T_9^{5/3} .
$
In Fig.\ 2 (dashed curve) the effect of electron shear is
maximized, using a very high $\mu_e/\mu_q=0.1$.
Surface rubbing due to the electron atmosphere being carried along by
the $r$-modes in the quark phase, scattering mainly on phonons in the
nuclear crust, corresponds to a viscous time scale
$
\tau_{\rm sr}\approx 1.42\times 10^8 {\rm s} T_9 (\nu/1{\rm kHz})^{-1/2}
$ 
for a crust with maximal density \cite{crust} (dash-dot curve).
For lower crust base density, the effect of surface rubbing is reduced further. 
Figure 2 shows $r$-mode instabilities in strange stars dominated by CFL
phase with
$\Delta =1$MeV. Much higher energy gaps (50--100 MeV) are expected in
recent studies of color superconductivity \cite{raj99}, but as seen from
Fig.~2, even a value of 1 MeV is incompatible with pulsar data, since
basically all rapid pulsars are located in the unstable regime,
and therefore should spin down by gravitational wave emission in a
matter of hours, c.f.\ $\tau_{\rm gw}$. Clearly in contradiction with
the facts.

At lower density, the high mass of the $s$-quark relative to $u$ and $d$
prevents creation of the CFL phase; instead two color states of $u$ and $d$
may pair, creating a 2-flavor color superconducting phase (2SC) that
introduces energy gaps for 4 out of 9 quark color-flavor states. If the
corresponding energy gap is of any significance,
the states with a gap can be safely ignored
compared to the remaining unpaired $s$-quarks and one color of $u$ and
$d$. This reduces the rate of the weak reaction $u+s\leftrightarrow d+u$
by a factor 1/9, increasing $\tau_{\rm bv}$ by
a factor 9. The strong scattering rates responsible for the shear
viscosity are reduced by $(5/9)^{1/3}$, thus increasing $\tau_{\rm sv}$ by
$(9/5)^{1/3}$. This expands the domain of $r$-mode instability as shown in
Fig.\ 3 to an extent where some rapid pulsars are in the unstable zone,
in disagreement with data. It is fair to say, though, that the
uncertainties and approximations involved may be large enough that
2SC quark matter stars may not be ruled out completely.

The $r$-mode instability thus provides
several interesting tests of the hypothesis of stable quark matter stars
(strange stars). If strange quark matter is absolutely stable, pulsars
would be expected to consist of quark matter. Data on pulsar rotation
properties are consistent with this if the quark matter is
non-superfluid (but only for strange stars with a thick crust or
completely bare
strange stars). The lack of observed very rapid pulsars may be due to the
$r$-mode instability, and rapid pulsars reaching the region of
instability will spin down in a characteristic manner, that can be
tested by observations. 

Strange stars in a color-flavor locked phase
are, in contrast, not permitted by pulsar data. Most rapid
pulsars would be $r$-mode unstable, and should spin down within hours,
which clearly they do not.
So if strange quark matter is stable, it may
be concluded that a CFL phase, and probably a 2SC
phase as well is not reached at
densities relevant in pulsars, i.e.\ up to a few times nuclear density.
These arguments do not rule out a color superconducting phase at such
densities if quark matter is only metastable, because then a
pulsar, even if it contains quark matter, will not have the separation of
the crust characteristic of a strange star. Thus, such a star is
susceptible to the full surface rubbing effect, and will not be $r$-mode
unstable to a similar degree. A detailed study of $r$-modes
in such hybrid stars would be interesting.

This work was supported in part by the Theoretical Astrophysics Center
under the Danish National Research Foundation.
I thank the referees and Krishna Rajagopal for comments on an earlier
version.

\begin{figure}
\epsfxsize=8.6truecm
\epsfbox{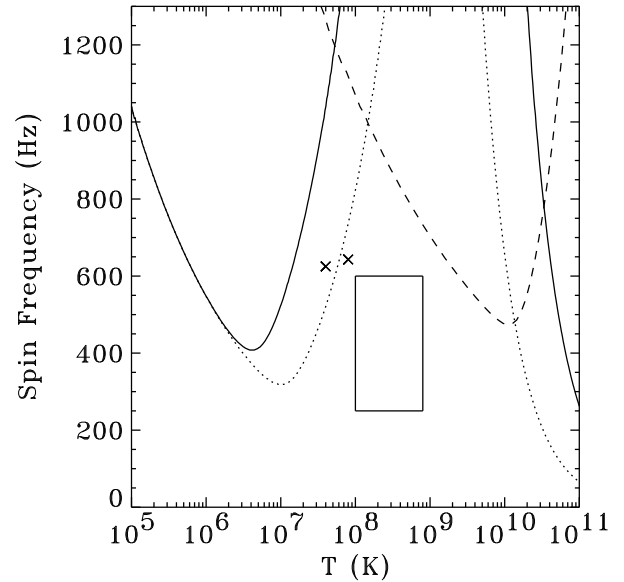}
\caption[]{Critical spin frequencies in Hz for strange stars
as functions of temperature.
Solid (dotted) curves assume $m_{100}=2$ (1).
The $r$-mode instability is active above the curves. The box shows
characteristic positions of LMXB's,
and crosses upper bounds on core temperatures for
the two fastest known pulsars. Dashed curve marks the $r$-mode
instability domain for neutron stars \cite{bilush99}.}
\end{figure}
\begin{figure}
\epsfxsize=8.6truecm
\epsfbox{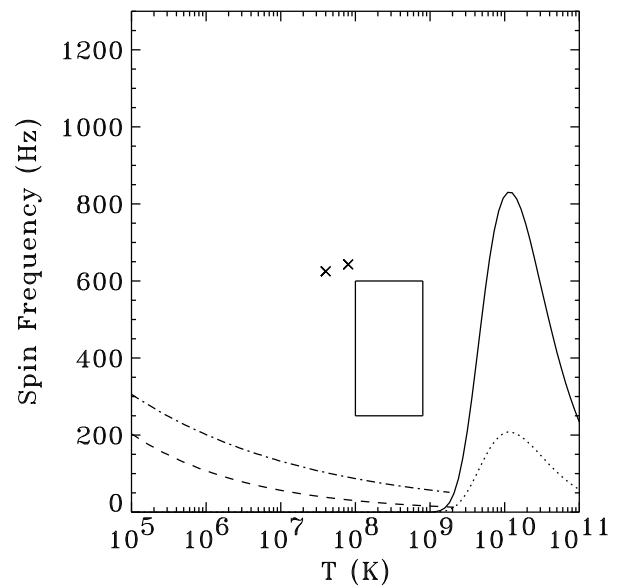}
\caption[]{As Fig.\ 1, but assuming quark matter in a
CFL phase with energy gap $\Delta=1$MeV.
Dashed curve results from electron shear; dash-dot curve from surface
rubbing for a maximal crust. Quark stars above the curves
are $r$-mode unstable, in clear disagreement with pulsar data.}
\end{figure}
\begin{figure}
\epsfxsize=8.6truecm
\epsfbox{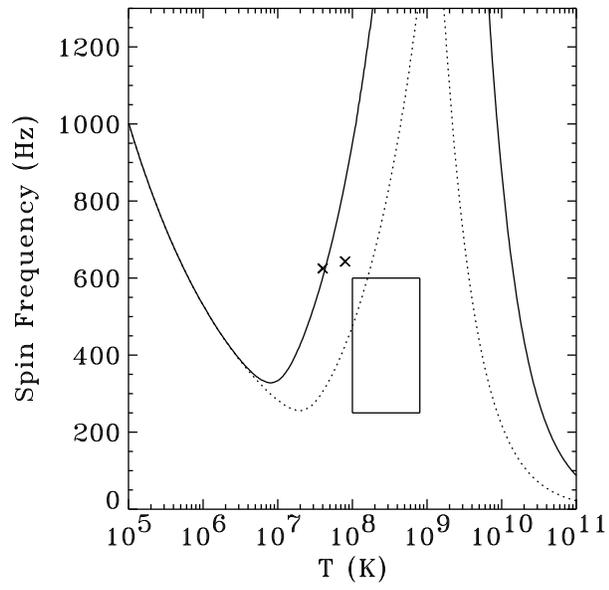}
\caption[]{As Fig.\ 1, but assuming a 2SC state.}
\end{figure}


\begin{references}
\bibitem{visc} Viscosity due to magnetic fields and vorticity
discussed in \cite{rezmaa99,relash99,holai99} would not change the main
conclusions of the present study.
\end{references}
\end{document}